\documentclass[aps,reprint,showpacs,amsmath,amssymb,pra,longbibliography,superscriptaddress]{revtex4-2}
\usepackage{amsmath,amssymb,latexsym,epsfig,graphics,epsf,siunitx}
\begin{document}
\title{Non-exponential relaxation without dynamic heterogeneity in van der Waals liquids above the melting point}


\author{Rolf Zeißler}
\affiliation{Institute for Condensed Matter Physics, Technical University of Darmstadt, D-
64289 Darmstadt, Germany}
\author{Niklas Pfeiffer}
\affiliation{Institute for Condensed Matter Physics, Technical University of Darmstadt, D-
64289 Darmstadt, Germany}
\author{Thomas Blochowicz}
\affiliation{Institute for Condensed Matter Physics, Technical University of Darmstadt, D-
64289 Darmstadt, Germany}

\begin{abstract}
We investigate the influence of dynamic heterogeneity on the spectral shape of structural relaxation in van der Waals liquids above the melting point by means of depolarized dynamic light scattering. To this end, we study optically anisotropic probe molecules both in the bulk and when diluted in an optically isotropic solvent. Strikingly, the relaxation shape of the probe molecules in dilution is indistinguishable from that of the pure liquid composed of the probe molecules. By contrast, when explicit dynamic heterogeneity is introduced, e.g., through internal degrees of freedom or a distribution of probe molecule sizes, the relaxation shape becomes sensitive to the solvent concentration. These findings indicate that dynamic heterogeneity has a negligible influence on the rotational dynamics of single component van der Waals liquids above the melting point, despite the pronounced non-exponential character of their relaxation shape.
\end{abstract}

\maketitle

\section{Introduction}

The prevailing view of rotational dynamics in molecular glass formers is that of homogeneous dynamics at high temperatures, which becomes increasingly heterogeneous as the liquid is cooled towards its glass transition temperature $T_{\mathrm{g}}$. This growing dynamic heterogeneity upon cooling is widely believed to manifest in the stretching of orientational correlation functions, as probed by various experimental techniques. The extent of relaxation stretching is commonly quantified by the stretching or width parameter $\beta$ appearing in different model functions, most notably the Kohlrausch-Williams-Watts (KWW) and the Cole-Davidson (CD) models. Within these models, a value of $\beta=1$ corresponds to a mono-exponential and values of $0<\beta<1$ indicate a stretched correlation decay.\cite{bohmer1998nanoscale,sillescu1999heterogeneity,angell2000relaxation,ediger2000spatially,richert2002heterogeneous,cavagna2009supercooled,richert2011experimental}

A large body of experimental and simulation work over the past decades has been dedicated to identifying the microscopic origin of relaxation stretching. In particular, three scenarios have been discussed: dynamics that is fully heterogeneous, partially heterogeneous, or fully homogeneous.
In the fully heterogeneous scenario, rotational correlation functions of different sub-ensembles within the system decay exponentially but with distinct relaxation times, resulting in a stretched decay of the ensemble-averaged correlations.
By contrast, the fully homogeneous scenario assumes that correlation decay is inherently stretched, with no variation in relaxation times among sub-ensembles.
Intermediate scenarios have also been considered, in which the correlation decay of sub-ensembles is inherently stretched, albeit to a lesser extent than that of the full ensemble, while an additional distribution of relaxation times across sub-ensembles gives rise to the observed stretching of the ensemble averaged correlation functions.
By employing experimental techniques capable of selecting and probing the dynamics of specific sub-ensembles, such as multi-dimensional NMR, dielectric hole burning, photobleaching experiments and fluorescence microscopy, compelling evidence for the purely heterogeneous scenario has been reported at temperatures close to $T_{\mathrm{g}}$.\cite{bohmer1998nature,cicerone1995molecules,schiener1996nonresonant,paeng2015ideal}

In the purely heterogeneous scenario detailed above, assuming no drastic change in the relaxation mechanism, the rotational dynamics of a liquid, as it is cooled from temperatures far above the melting point $T_{\mathrm{m}}$ towards $T_{\mathrm{g}}$, would be expected to show a monotonic decrease in $\beta$, starting from $\beta_{\mathrm{hi}}\approx 1$ in the high-temperature limit, alongside an increase of a corresponding length scale \cite{Kim2022a, Johari2012a}. While this expectation is rarely questioned explicitly, it is in many instances contrasted by experimental results.

In several depolarized dynamic light scattering (DDLS) studies, Rössler et al.\ reported virtually temperature-independent relaxation stretching, with CD width parameters ranging from 0.32 to 0.8, observed from temperatures below $T_{\mathrm{m}}$ up to, in some cases, the boiling point $T_{\mathrm{b}}$ of a variety of molecular glass formers. The authors emphasize that a crossover to "simple liquid" dynamics, characterized by one-step correlation decay and the disappearance of relaxation stretching, is only observed at the highest temperatures, close to $T_{\mathrm{b}}$.
Furthermore, this crossover does not manifest as a gradual narrowing of the correlation functions, but rather as a merging with vibrational dynamics, leaving only the long time decay of rotational correlations accessible, which naturally matches that of a simple exponential.\cite{petzold2013evolution,schmidtke2014relaxation,rossler2025relaxation}

However, this observation is not limited to light scattering. Lunkenheimer et al. performed extensive broadband dielectric spectroscopy (BDS) studies on benzophenone, glycerol and propylene carbonate, covering timescales of molecular rotation from picoseconds to several thousand seconds. They convincingly demonstrated that CD width parameters, obtained by fitting the dielectric loss peak, although increasing with temperature, do not reach unity but saturate at values $\beta_{\mathrm{hi}}<1$, that are characteristic of each liquid under study.\cite{lunkenheimer2007benzophenone,lunkenheimer2000glassy}

Besides experimental evidence there exist also simulation studies that report this behavior for molecular glass formers. Eastwood et al. have shown for an atomistic model of ortho-terphenyl that relaxation stretching of correlation functions reflecting molecular reorientation decreases upon increasing the temperature from well below the melting point and then becomes weakly temperature dependent or even constant at high temperatures, with values of the stretching parameter $\beta_{\mathrm{hi}}<1$\cite{eastwood2013rotational}. Similar results have been obtained in a first principles study on the glass former toluene, however, a straight forward link between a measure of spatial dynamic heterogeneity and relaxation stretching remains elusive \cite{pabst2025glassy}.

Thus, on the one hand there exists a temperature regime in which relaxation stretching remains constant, while still being far from mono-exponential, spanning temperatures from some value above $T_{\mathrm{g}}$ to temperatures well above $T_{\mathrm{m}}$, and on the other hand experimental evidence shows that relaxation stretching is connected to dynamic heterogeneity near $T_{\mathrm{g}}$. Therefore, dynamic heterogeneity needs to be investigated in this high-temperature regime. Specifically, one may ask if the observed relaxation stretching in this regime is inherent to the local rotational correlation functions, or whether some virtually temperature independent form of dynamic heterogeneity persists even at temperatures well above $T_{\mathrm{m}}$.

In the present work, we scrutinize these conjectures by performing DDLS on different optically anisotropic probe molecules above their respective melting points, both in the bulk and in dilution with an optically isotropic solvent. All probe molecules show an asymmetrically broadened relaxation peak in their susceptibility spectra, indicating non-exponential relaxation in their bulk form. We propose that if this non-exponentiality arises from dynamic heterogeneity, the relaxation shape should change upon dilution with solvent molecules. However, our results show that this is not the case, suggesting that the extent of relaxation stretching at these temperatures reflects the intrinsic properties of the individual molecules, such as molecular shape and internal flexibility. Thus, given the heterogeneous scenario previously identified close to the glass transition is correct, a drastic change of the relaxation mechanism has to occur between $T_{\mathrm{g}}$ and $T_{\mathrm{m}}$.

\section{Experimental Details}

\subsection{Samples}

\begin{table*}
\centering
\caption{Table of investigated substances.}
     \begin{tabular}{llllllll}\hline\hline
       substance  & abbreviation & supplier & chemical purity (\SI{}{\percent}) & $T_{\mathrm{m}}$ (K) & $M$ (g/mol) \\\hline
       diethyl phthalate&DEP&Sigma Aldrich&$\geq99.5$&270&222.24\\
       tributyl phosphate&TBP&Sigma Aldrich&$\geq99$&194&266.32\\
       n-tridecane&C13&Alfa Aesar&$\geq98$&268&184.37\\
       n-tetradecane&C14&Thermo Scientific&$\geq99$&279&198.39\\
       n-heptane&C7&Thermo Scientific&$\geq99$&183&100.21\\
       1-phenyltridecane&P13&TCI&$\geq99$&283&260.46\\
       carbon tetrachloride&CCl$_4$&Sigma Aldrich&$\geq99.5$&250&153.82\\
       \\\hline\hline
     \end{tabular}
     \label{tbl:table1}
\end{table*}

Tbl.\,\ref{tbl:table1} contains the abbreviations, suppliers, chemical purities, melting temperatures and molecular weights of the liquids used in this study.
The liquids were used without further purification and mixtures were prepared by weighing the different components to obtain the desired molar fraction with an uncertainty of 0.1\,mol\SI{}{\percent}.

CCl$_4$ was chosen as a solvent since it is optically isotropic and should therefore contribute little to the total scattered intensity in a DDLS experiment.\cite{cummins1996origin}

DEP, TBP and the linear n-alkanes were chosen as probe molecules for several reasons.

Firstly, they exhibit very different relaxation shapes in bulk. As demonstrated in the SI of this work, the relaxation shapes of these liquids are temperature independent in moderate temperature ranges above their respective melting point, as it was also demonstrated for a number of molecular liquids in the literature\cite{schmidtke2014relaxation}, with different extent of asymmetric broadening. Fits to the relaxation peak region by the Cole-Davidson model yield width parameters of $\beta_{\mathrm{CD}}=0.74$ for DEP, $\beta_{\mathrm{CD}}=0.53$ for TBP and $\beta_{\mathrm{CD}}=0.83$ for C13. Thus, for all three liquids rotational dynamics has a pronounced non-exponential character that is essentially temperature independent in the investigated temperature range.

Secondly, the peak frequencies of the relaxation peak in bulk spectra are very similar and close to the low frequency limit of the experimentally accessible frequency range at 300\,K, which results in a large separation of the relaxation peak from the vibrational part of the spectra, giving the best possible starting point for this investigation.
This fact also allowed us to performed all measurements at a temperature of 300\,K, giving the most controlled and reproducible conditions, since no optical cryostat needs to be involved which can be subject to condensation on windows, stress birefringence etc..
Conveniently, 300\,K lies moderately above the melting point of the chosen liquids, essentially making sure that the liquids are not studied in completely different dynamic regimes.

Mixtures of linear n-alkanes of different alkyl chain length allow to study the effect of dilution on probe molecules with a molecular size distribution. Here, the idea is to introduce an explicit form of dynamic heterogeneity or distribution of relaxation times through probe molecules with different molecular weights and sizes and, therefore, different mobilities and to investigate the influence dilution with CCl$_4$ has on such a system compared to a single component liquid.

In addition to these rather simple probe molecules, we chose P13 as a probe molecule with a strong influence of internal degrees of freedom on the relaxation shape. As demonstrated in refs.\,\citenum{zeissler2023influence,zeissler2025relation,zeissler2025role}, molecules consisting of an aromatic ring attached to a linear alkyl chain exhibit a bimodal relaxation peak due to internal rotation of the aromatic ring, which opens up another avenue to study the effect of dilution on systems with an explicit form of dynamic heterogeneity, in this case, of intramolecular nature.

\subsection{Depolarized dynamic light scattering}

Depolarized dynamic light scattering (DDLS) spectra were obtained by a multipass tandem Fabry-Perot interferometer TFP-1 by JRS Scientific Instruments ($\nu\sim200$\,MHz-200\,GHz) and a U1000 Jobin Yvon double monochromator ($\nu\sim200$\,GHz-5\,THz).

In both experiments the depolarized component of the spectral density $I(\nu)$ is measured and the imaginary part of the DDLS susceptibility $\chi^{\prime\prime}(\nu)$ is then calculated via the fluctuation dissipation theorem:
\begin{equation}
\chi^{\prime\prime}(\nu)=\frac{I(\nu)}{n(\nu,T)+1}
\label{eq:fdt}
\end{equation}
where $n(\nu,T)=(\mathrm{exp}(h\nu/k_{\mathrm{B}}T)-1)^{-1}$ is the Bose temperature factor.

Correct relative values of $\chi^{\prime\prime}$ are obtained by repeated measurement of a standard, in this case CCl4, during every measurement series.

The integrated depolarized scattering intensity $I_{\mathrm{tot}}$ can be obtained either through integrating directly $I(\nu)$ or integrating $\chi^{\prime\prime}(\nu)\cdot (n(\nu,T)+1)$. The latter approach was used in this work.

\section{Experimental Results}

\subsection{Dilution of single component van der Waals liquids}

\begin{figure}[ht!]
  \includegraphics[width=0.5\textwidth]{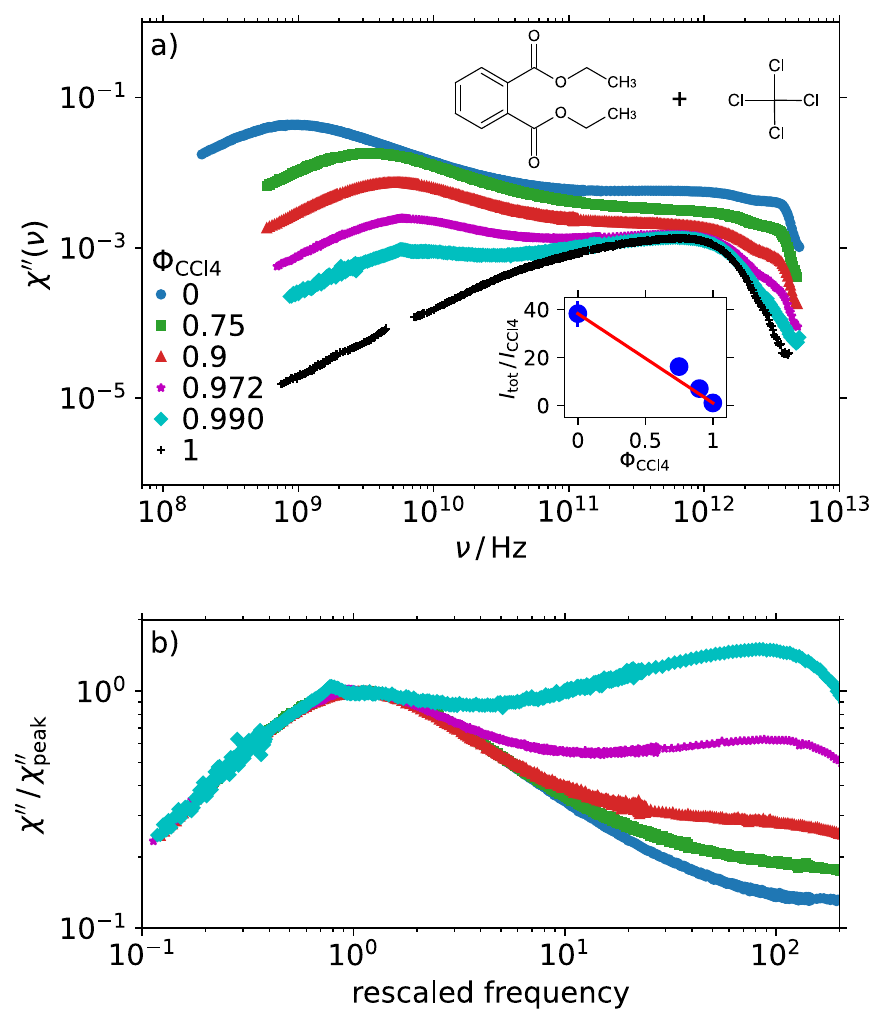}
  \caption{a) DDLS spectra of DEP, CCl$_4$ and mixtures of both liquids at a molar fraction of CCl$_4$ of $\Phi_{\mathrm{CCl4}}$. The inset shows the integrated depolarized scattering intensity $I_{\mathrm{tot}}$ normalized by the intensity scattered from pure CCl$_4$. The red line represents an ideal mixing law for the scattered intensity of the mixture. b) DDLS spectra of (a) normalized to the amplitude of the relaxation peak and shifted in frequency to collapse the low frequency flanks.}
  \label{fig:DEP_dilution}
\end{figure}

Fig.\,\ref{fig:DEP_dilution} shows DDLS spectra of pure DEP and CCl$_4$ and mixtures of both at molar fractions $\Phi_{\mathrm{CCl4}}$ of 0.75, 0.907, 0.972 and 0.99 at 300\,K. Part (a) of the figure shows the non-normalized spectra, whereas in part (b) of the figure the spectra are presented normalized to the maximum amplitude of the relaxation peak and shifted in frequency to collapse the low frequency flanks.

Noticeably, the spectrum of CCl$_4$ shows much faster dynamics with the relaxation peak being visible only as a weak shoulder on the low frequency side of the vibrational part of the spectrum, as already reported in the literature \cite{cummins1996origin}. 
We mention that while for molecules with sufficient optical anisotropy molecular reorientation is directly monitored by depolarized light scattering, optically isotropic solvents, like CCl$_4$, scatter by the dipole-induced-dipole (DID) mechanism, which is not as straight forward to interpret \cite{pabst2022intensity, cummins1996origin}. But ab-initio simulations recently demonstrated that also the DID part of the DDLS spectra, even if dominant, still reflects structural relaxation \cite{pabst2025water}, so that the CCl$_4$ spectrum is not qualitatively different, it just shows faster relaxation with a significantly lower amplitude.

\begin{figure*}[ht!]
  \includegraphics[width=\textwidth]{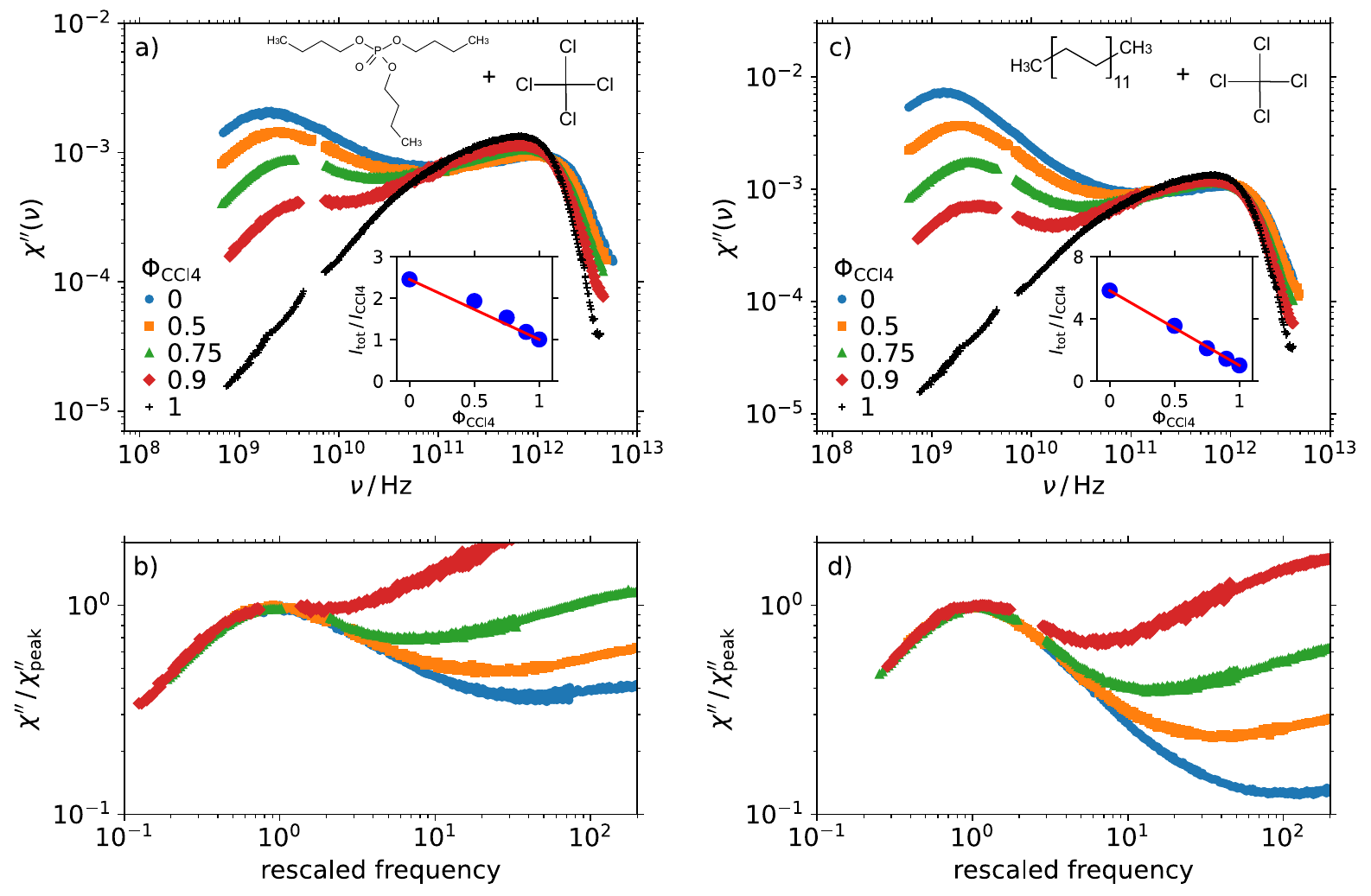}
  \caption{Analogous analysis to Fig.\,\ref{fig:DEP_dilution} for probe molecules TBP (a-b) and C13 (c-d).}
  \label{fig:massive_multipanel_figure}
\end{figure*}
For DEP as well as the mixtures of DEP and CCl$_4$ a pronounced relaxation peak is visible that shifts to higher frequencies with increasing CCl$_4$ concentration.
As shown in the inset of part a) of the figure, the integrated intensity $I_{\mathrm{tot}}$ of pure DEP is 38 times as high as that of CCl$_4$ and even at $\Phi_{\mathrm{CCl4}}=0.907$ the integrated intensity is still 7 times as high as that of pure CCl$_4$. It is therefore clear that, even in moderate dilution, the relaxation spectra are dominated by the contributions of the optically anisotropic DEP molecules.
Strikingly, the spectral shape of the relaxation peak is unaltered relative to that of pure DEP up to $\Phi_{\mathrm{CCl4}}=0.907$. The only change that can be observed is a decrease in amplitude of the relaxation peak relative to the vibrational part of the spectrum. For higher concentrations $\Phi_{\mathrm{CCl4}}>0.907$ the relaxation peak is not well enough resolved to allow for a straight forward conclusion based on the raw data.

Fig.\,\ref{fig:massive_multipanel_figure} shows a similar analysis as presented in Fig.\,\ref{fig:DEP_dilution} for the liquids TBP (a-b) and C13 (c-d) at 300\,K. Part (a) and (b) of the figure show the non-normalized DDLS spectra with the insets again showing $I_{\mathrm{tot}}\,/\,I_{\mathrm{CCl4}}$ in dependence on $\Phi_{\mathrm{CCl4}}$.

Part (c) and (d) show the DDLS spectra normalized to the maximum amplitude of the relaxation peak and shifted in frequency. Compared to DEP these liquids show significantly lower depolarized scattering intensity and the relaxation peaks are less resolved.
Up to $\Phi_{\mathrm{CCl4}}=0.75$ the same observation as for dilution of DEP is readily apparent: the shape of the relaxation peak is not significantly altered by the addition of CCl$_4$.

For $\Phi_{\mathrm{CCl4}}>0.75$ the relaxation peak again is not resolved enough anymore to draw this conclusion from simple comparison of the spectra.

\begin{figure}[ht!]
  \includegraphics[width=0.5\textwidth]{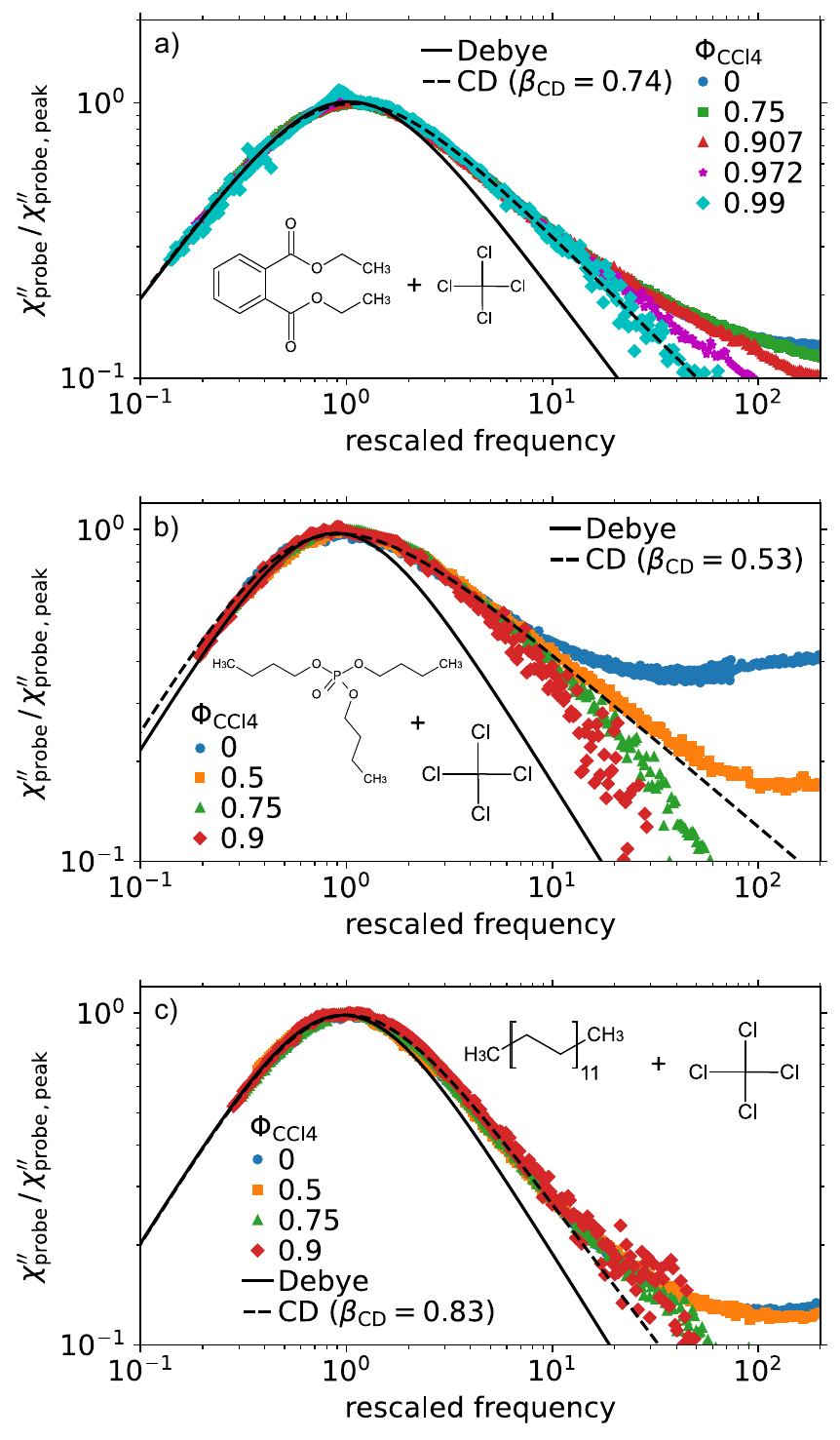}
  \caption{DDLS spectra after subtraction of the CCl$_4$ contribution according to eq.\,\ref{eq:subtraction} for DEP (a), TBP (b) and C13 (c).}
  \label{fig:dilution_with_subtraction}
\end{figure}

To deal with this problem, we subtracted an estimation of the CCl$_4$ contribution from the relaxation spectra under assumption of an ideal mixing law:
\begin{equation}
    \chi^{\prime\prime}_{probe}(\nu)=\chi^{\prime\prime}_{total}(\nu)-\Phi_{\mathrm{CCl4}}\cdot\chi^{\prime\prime}_{CCl4}(\nu)\,.
\label{eq:subtraction}
\end{equation}
Here $\Phi$ is the molar fraction of CCl$_4$, $\chi^{\prime\prime}_{total}(\nu)$ the total DDLS spectrum, $\chi^{\prime\prime}_{CCl4}(\nu)$ the DDLS spectrum of pure CCl$_4$ and $\chi^{\prime\prime}_{probe}(\nu)$ the estimated contribution of the probe molecules to the total DDLS spectrum. For this subtraction method to work, the contribution of CCl$_4$ in the mixtures has to be largely unaltered with respect to its bulk spectrum, apart from being scaled by concentration. This assumption is not obvious, however, is justified by the results of this subtraction procedure presented in Fig.\,\ref{fig:dilution_with_subtraction}. We suggest that this is due to the fact that the CCl$_4$ DDLS spectrum has an incredibly weak relaxation contribution and is mostly dominated by high frequency vibrational contributions which are known to be less sensitive to, e.g., viscosity changes.

For all probe molecules and CCl$_4$ concentrations the diluted spectra collapse onto the relaxation peak of the pure liquid after subtraction of the CCl$_4$ contribution, even at CCl$_4$ concentrations as high as $\Phi_{\mathrm{CCl4}}=0.99$ for DEP. Black solid and dashed curves represent the susceptibilities of the Debye- and Cole-Davidson models respectively to highlight again the pronounced non-exponential character of probe molecule rotation up to the highest molar fractions of CCl$_4$.

\subsection{Dilution of probe-molecules with internal degrees of freedom}

\begin{figure}[ht!]
  \includegraphics[width=0.5\textwidth]{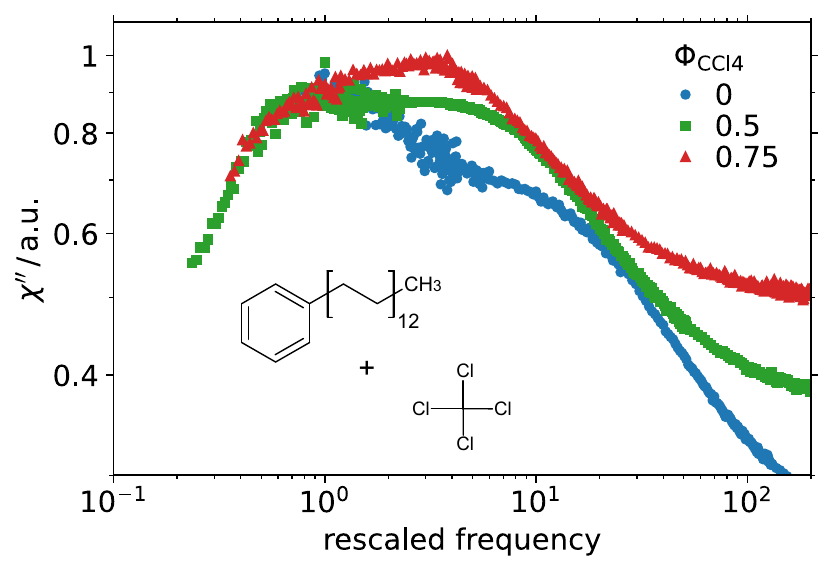}
  \caption{DDLS spectra of P13 and mixtures of P13 and CCl$_4$ at 300\,K. The spectra have been shifted in frequency and amplitude for comparison of the peak shapes.}
  \label{fig:P13_dilution}
\end{figure}

Fig.\,\ref{fig:P13_dilution} shows the DDLS spectrum of P13, consisting of a phenyl ring attached to a linear alkyl chain of length $n=13$, and mixtures of P13 and CCl$_4$ at 300\,K shifted in frequency and amplitude. As demonstrated in refs.\,\citenum{zeissler2023influence, zeissler2025relation, zeissler2025role}, molecules consisting of an aromatic ring attached to an alkyl chain exhibit bimodal relaxation peaks due to internal rotation of the aromatic ring. The fast contribution to the bimodal relaxation peak has been assigned to ring rotation, while the slow contribution is connected to rotation of the entire molecule. This can be interpreted as a case where dynamic heterogeneity, in this case \emph{intramolecular dynamic heterogeneity}, clearly influences the relaxation shape.

Interestingly, dilution by CCl$_4$ has a significant impact on the relaxation shape in this case. The separation of the two contributions decreases with increasing CCl$_4$ contribution together with a possible change of the relative amplitude of the two contributions. However, disentangling the influence of dynamic separation and relative amplitude requires use of an explicit fit model, which is out of the scope of the current work.

\subsection{Dilution of probe-molecules with a distribution of molecular weights}

\begin{figure}[ht!]
  \includegraphics[width=0.5\textwidth]{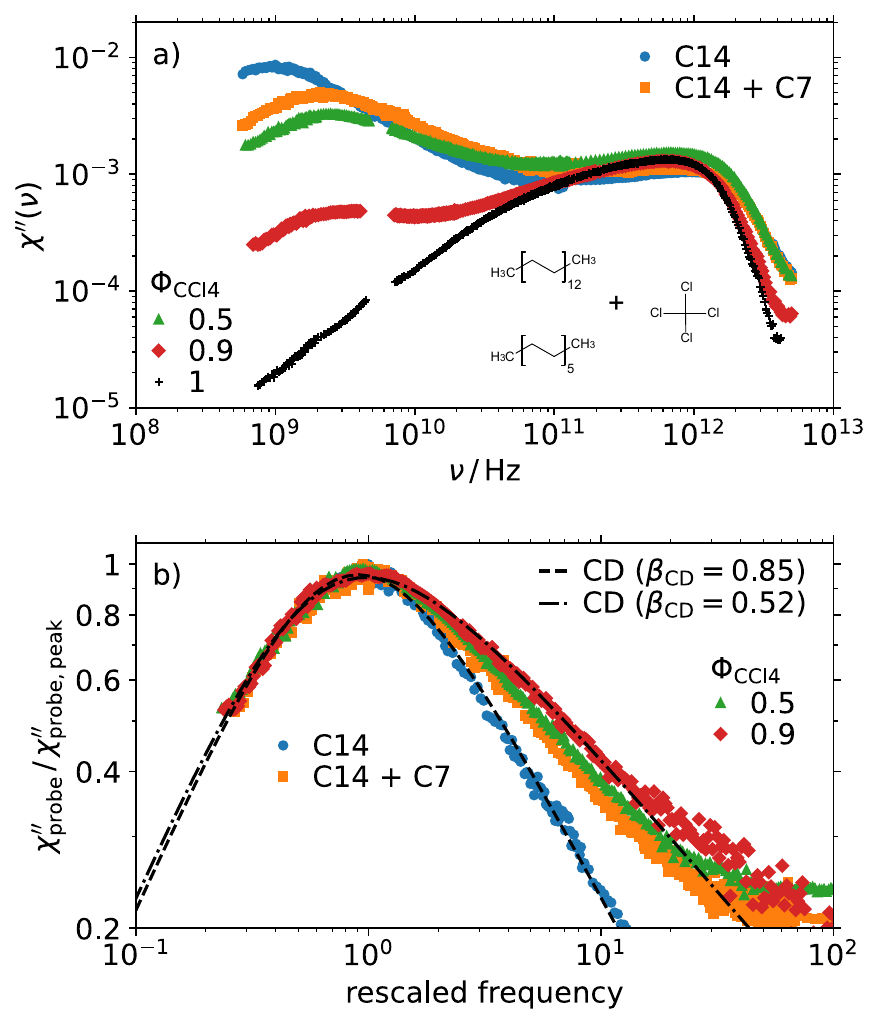}
  \caption{a) DDLS spectra of C14, binary mixture of C14 and C7 at a molar fraction of both constituents of 0.5 and tertiary mixtures of C14, C7 and CCl$_4$ (see text). b) Normalized DDLS spectra after subtraction of the CCl$_4$ contribution according to eq.\,\ref{eq:subtraction}.}
  \label{fig:C14_C7_dilution}
\end{figure}

Fig.\,\ref{fig:C14_C7_dilution} (a) shows non-normalized DDLS spectra of n-tetradecane (C14), a binary mixture of C14 and n-heptane (C7) at a molar fraction of both constituents of 0.5, as well as tertiary mixtures of C14, C7 and CCl$_4$. In this case $\Phi_{\mathrm{CCl4}}=0.5$ corresponds to molar fractions of C14 and C7 of 0.25 respectively and $\Phi_{\mathrm{CCl4}}=0.9$ to molar fractions of each alkane of 0.05. C14 and C7 differ in their molecular weight $m_{\mathrm{w}}$ by a factor of $\sim2$ with $m_{\mathrm{w}}=$198.39\,g/mol for C14 and $m_{\mathrm{w}}=$100.21\,g/mol.

Part (b) of the figure shows the normalized spectra after subtraction of the CCl$_4$ contribution according to eq.\,\ref{eq:subtraction}. The relaxation peak of the binary mixture of C14 and C7 is significantly broadened with respect to the relaxation peak in the spectrum of pure C14, likely due to different mobility of the C14 and C7 molecules in the mixture. This essentially represents a form of dynamic heterogeneity through distribution of molecular weights or sizes and therefore molecular mobilities. Strikingly, in contrast to the dilutions of probe molecules with uniform molecular weight, the relaxation peaks in the tertiary mixtures of C14, C7 and CCl$_4$ broaden even further with increasing CCl$_4$ concentration. CD fits of the peak region yield width parameters $\beta_{\mathrm{CD}}$ of 0.85 for pure C14, 0.68 for the binary mixture of C14 and C7 and 0.62 at $\Phi_{\mathrm{CCl4}}=0.5$ and 0.52 at $\Phi_{\mathrm{CCl4}}=0.9$ for the tertiary mixtures after subtraction of the CCl$_4$ contribution.

\section{Discussion}

Under the assumption that the spectral shape of the relaxation peak contains information on dynamic heterogeneity and the underlying physical distribution of relaxation times, dilution should alter the shape, since it significantly changes the local environment experienced by the probe molecules. At low probe molecule concentrations this resembles earlier works where dielectric probe molecules were, e.g., used to investigate properties of non-polar solvents by dielectric spectroscopy \cite{Williams1971b,Johari1972a}.

Regarding the question if a probe is sensitive to dynamic heterogeneity of the host system, three scenarios can be distinguished:

(i) If the probe molecules are significantly larger and the probe dynamics much slower than the dynamics of the host, the probe molecules experience an averaged quasi-homogeneous environment and are unaffected by heterogeneity. In this case, the relaxation shape of the probe molecules should be close to mono-exponential, exhibiting a symmetric Debye-like relaxation peak in the susceptibility spectra, as was, e.g., demonstrated by optical probes in ortho-terphenyl \cite{cicerone1995molecules, paeng2015ideal, Paeng2018a}.

(ii) If the solvent is heterogeneous on the time- and/or length scale of probe molecule rotation, the latter will effectively probe the heterogeneity of the solvent \cite{Mackowiak2011a, Leone2013a,paeng2015ideal}. As a result, the relaxation shapes of all three molecular probes, DEP, TBP and C13 in CCl$_4$, should differ from each of the respective bulk spectra and ideally the spectra of all probe molecules should become identical at sufficiently high CCl$_4$ concentrations, representing the extent of dynamic heterogeneity in the solvent.

(iii) If the dynamics of the probe molecules in bulk, as well as of the solvent in the mixtures, is homogeneous on the timescale of probe molecule rotation, but the rotational correlation decay of the probe molecules is inherently stretched, i.e., exhibits a non-exponential character due to some intrinsic property of the molecule, no change of the relaxation shape upon dilution is expected.

From the presented results scenario (iii) seems to hold for the three van der Waals liquids, DEP, TBP and C13.

We note that in previous studies of molecular probes close to the glass transition of a host solvent the hydrodynamic limit, i.e. a Debye-like relaxation shape was approached if ratios of molecular mass of probe $m_p$ and solvent $m_s$ exceeded $m_p/m_s \approx 1.3$ and the probe relaxation time was slower than the solvent one by a factor of 20 or more \cite{Wang2004b,Yang2002a}. 
Comparing to our own results, the mentioned mass ratio is exceeded by DEP and TBP (cf.\ Tab.\ \ref{tbl:table1}) and also the estimated time scale separation from Figs. 1 and 2 would lead to expecting scenario (i) while actually scenario (iii) is observed, i.e., no change in the respective Cole-Davidson parameters of DEP and TBP, namely $\beta_{\mathrm{CD}} = 0.74$ and $\beta_{\mathrm{CD}}=0.53$, in the bulk and in mixtures of any concentration is observed. Of course, the main difference of the above results and the mentioned earlier observations is, besides the applied experimental technique, the temperature range, which is above the melting point for the presented data and not in the deeply supercooled liquid. Thus, we take this result as an indication that relaxation mechanisms differ drastically between the deeply supercooled and the liquid regime.

As mentioned in the introduction, there is compelling experimental evidence for the purely heterogeneous scenario close to the glass transition, i.e., relaxation stretching occurs  due to distinct relaxation times of different sub-ensembles with mono-exponential relaxation functions. This indirectly implies a continuous return to mono-exponential relaxation when dynamic heterogeneity diminishes at high temperatures. The fact that the spectral shape of the investigated probe molecules is not altered while the molecular environment for the probe molecules changes from bulk to low concentration suggests that there is no influence of dynamic heterogeneity on the relaxation shape at the studied temperatures. Moreover, the shape is significantly non-exponential and independent of temperature in a considerable range of timescales. Therefore it seems likely that there is a change of mechanism in structural relaxation somewhere between $T_{\mathrm{g}}$ and $T_{\mathrm{m}}$ resulting in a transition from detectable heterogeneous relaxation in the deeply supercooled regime to effectively homogeneous dynamics with inherent relaxation stretching in the liquid regime. 

Evidence for a change in transport mechanism and a crossover from liquid to glass-like dynamics at a certain temperature or timescale between $T_{\mathrm{g}}$ and $T_{\mathrm{m}}$ has been discussed since long \cite{Goldstein1969a} and is often thought to be related to the critical temperature of mode-coupling theory (MCT) \cite{Goetze1992a}, although its exact nature remains elusive \cite{novikov2003universality}.
Notably, besides the possible change in the origin of relaxation stretching postulated in the present work, dynamics of molecular glass formers exhibits several drastic changes in this temperature regime, such as splitting of $\alpha$- and $\beta$ relaxation processes \cite{rossler1990indications}, decoupling of rotational and translational dynamics \cite{rossler1990indications, chang1997heterogeneity,swallen2003self} and diminishing influence of internal degrees of freedom and intrinsic molecular properties on structural relaxation \cite{zeissler2025role}.

Another consequence of the present observations concerns the influence of orientational cross-correlations on the relaxation shape. In contrast to dielectric spectroscopy, where orientational cross-correlations of molecular dipoles contribute significantly to the relaxation spectra \cite{bohmer2025spectral}, there is little evidence so far that this is the case in DDLS, to a comparable extent, likely due to the different ranks of the underlying orientational correlation functions probed by the two techniques \cite{henot2023orientational}. In Ref.\,\citenum{pabst2020dipole} it was shown that the contribution of orientational cross-correlations in dielectric loss spectra of TBP can be suppressed by dilution with a nonpolar solvent, resulting in a match of the relaxation shapes probed by BDS and DDLS at a solvent molar fraction of 0.71, with no further change of the relaxation shape upon increasing the solvent concentration. The authors suggest that this suppression of cross-correlations results from the non-polar solvent molecules essentially shielding the dipole-dipole interactions of the polar probe molecules, which are the source of dipolar cross-correlations. This leads us to believe that the independence of the relaxation shape on CCl$_4$ concentration observed in the present work strongly supports the notion that the influence of orientational cross-correlations on the DDLS spectra of the three studied van der Waals liquids above $T_{\mathrm{m}}$ indeed is negligible. 

In the case of P13, a liquid with intramolecular dynamic heterogeneity, the relaxation shape is significantly altered upon dilution, i.e., the dynamic separation of the two contributions in the bimodal relaxation peak decreases, which may be due to the plasticizing effect of CCl$_4$ molecules, which makes the cooperative motion of the whole molecule approach the more local ring rotation \cite{zeissler2025relation}. For the n-alkane mixtures, on the other hand, the relaxation peak broadens upon addition of CCl$_4$, likely because on sufficient dilution by the small spherical CCl$_4$ molecules, the interactions between alkane chains of different length are shielded, effectively allowing a larger timescale separation of shorter and longer chains. Altogether, this simply demonstrates that any imposed dynamic heterogeneity immediately shows up as a change in relaxation shape and is sensitive to the solvent concentration. This behavior, although expected, contrasts the observations made for the single component van der Waals liquids DEP, TBP and C13, further confirming the inherent nature of relaxation stretching in these cases.

\section{Conclusions}

We demonstrated by depolarized dynamic light scattering on the van der Waals liquids diethyl phthalate, tributyl phosphate and n-tridecane, that the non-exponential relaxation shape of these molecules above their respective melting point is not affected by dilution with an optically isotropic solvent, essentially contradicting any expectation that would arise under the assumption that the non-exponentiality arises from dynamic heterogeneity at these temperatures.

Furthermore, these results indicate that there is no significant influence of orientational cross-correlations on the relaxation shape in this temperature regime, since cross-correlations would be expected to be suppressed upon dilution with a non-polar solvent, leading to a change of the relaxation shape.

For probe molecules with a size distribution or internal degrees of freedom the situation is quite different. Here, dilution has a significant effect on the relaxation shape, suggesting that intermolecular interactions play an important role in the extent of relaxation stretching or distribution of relaxation times in these cases.

The presented results contradict the widespread assumption that asymmetrically broadened relaxation peaks are necessarily a symptom of dynamic heterogeneity, which is supposed to vanish at sufficiently high temperatures, resulting in the absence of relaxation stretching. Instead the results suggest that the asymmetrically broadened shape of relaxation peaks observed above the melting point of one component liquids purely reflects specific properties of single molecules.

Assuming that the purely heterogeneous scenario holds close to $T_{\mathrm{g}}$, we suggest that our findings indicate a change in the mechanism of structural relaxation somewhere between $T_{\mathrm{g}}$ and $T_{\mathrm{m}}$ resulting in a transition to homogeneous dynamics with inherent relaxation stretching at high temperatures, a conjecture which we will pursue further in future studies.

\begin{acknowledgments}
Financial support by the Deutsche Forschungsgemeinschaft under grant no.\ 1192/3 is gratefully acknowledged.
\end{acknowledgments}

\section*{Author Declarations}

\subsection*{Conflicts of Interest}

The authors have no conflicts to disclose.

\subsection*{Author Contributions}

\textbf{Rolf Zeißler:} Writing -- original draft (lead), Visualization (lead), Investigation (lead), Formal analysis (lead), Data curation (lead), Conceptualization (equal). \textbf{Niklas Pfeiffer:} Writing -- review \& editing (supporting), Investigation (supporting), Formal analysis (supporting). \textbf{Thomas Blochowicz:} Writing -- review \& editing (lead), Supervision (lead), Funding acquisition (lead), Conceptualization (equal).

\bibliography{dilution.bib}

\end{document}